\documentclass{article}

\usepackage{bioarxiv}
\usepackage[utf8]{inputenc}
\usepackage[T1]{fontenc}
\usepackage{booktabs}
\usepackage{microtype}
\usepackage{graphicx}
\usepackage{amsmath}
\usepackage{newtxtext}
\usepackage{newtxmath}
\usepackage{tabularx}
\usepackage{setspace}
\usepackage{hyperref}

\makeatletter
\renewcommand{\section}{%
  \@startsection{section}{1}{\z@}%
  {-2.0ex \@plus -0.5ex \@minus -0.2ex}%
  {1.5ex \@plus 0.3ex \@minus 0.2ex}%
  {\normalfont\large\bfseries\raggedright}%
}
\makeatother

\title{Reading Between the Reads: A Decade On, Revisiting Sequencing Protocol Annotation in the Sequence Read Archive}

\author{
  Dr Jamie J.\ Alnasir\\
  Independent Researcher, London, United Kingdom\\
  \textit{Corresponding author:} \texttt{jamie.a8@gmail.com}
}

\begin{document}
\maketitle

\begin{abstract}

\noindent\textbf{Background:}
Experimental metadata are essential for interpreting and reusing publicly archived sequencing data. In 2015, Alnasir and Shanahan investigated the annotation of three library preparation steps --- fragmentation, adapter ligation and enrichment --- in the Sequence Read Archive (SRA). A decade later, the present study revisits these measurements using the original SQL keyword definitions against a September 2026 NCBI metadata snapshot reconstructed using the historical SRAdb relational schema.

\noindent\textbf{Results:}
The contemporary database contained 41,593,302 experiments, compared with 419,620 in the original study. In \texttt{library\_construction\_protocol}, the proportion matching all three historical keyword groups fell from 4.06\% in 2013 to 1.42\% in 2026; 67.65\% of contemporary experiments had a NULL value in this field. Searching \texttt{design\_description} as well identified 1,197,367 experiments (2.88\%) satisfying the three-step criterion across the two fields. Of these, 605,841 (50.60\%) would be missed by a protocol-field-only search. Among the 605,757 experiments matching all three groups exclusively in the design description, 604,116 (99.73\%) had a NULL protocol field. Combined-field matching rates varied by sequencing strategy, from 0.26\% for \texttt{AMPLICON} to 7.69\% for \texttt{RNA-Seq}. Historical keyword classifications agreed for all 150 purposively sampled experiments independently checked against source XML.

\noindent\textbf{Conclusions:}
Under the fixed historical keyword definitions, searching only the designated SRA library construction field misses just over half of the experiments meeting the three-step criterion across the two examined fields. This has direct implications for computational retrieval of archived preparation-related terminology. Historical changes in matching rates must nevertheless be interpreted in light of differences in sequencing strategies and library preparation methods. Keyword occurrence does not establish that a laboratory procedure was performed or adequately documented.

\end{abstract}


\section{Introduction}

Sequencing data deposited in public archives are frequently reanalysed as computational methods improve or new research questions arise. They are also used in comparative studies involving samples generated in different laboratories. Such work depends on reliable experimental metadata. The sequence reads provide little information about the laboratory procedures used to prepare the sequencing library, although differences in sample handling and library preparation can introduce systematic biases in the resulting data \cite{ross2013bias,alnasir2015annotation}.

Several steps in library preparation may influence the representation of sequences in the final dataset. Fragmentation affects the distribution of molecules entering the library; adapter attachment prepares those molecules for subsequent processing; and amplification or enrichment may favour particular sequences. The steps performed depend on the assay and sequencing platform. Some methods omit stages used in conventional workflows, whereas others combine them. A missing description therefore gives no reliable indication of whether a step was performed. Similarly, mentioning a procedure without describing its conditions may be insufficient for another researcher to assess its effects. These distinctions are relevant when archived data are used in studies of sequencing bias \cite{alnasir2015annotation,kozarewa2009amplification}.

The Sequence Read Archive (SRA), part of the International Nucleotide Sequence Database Collaboration, stores raw sequencing data with associated experimental metadata \cite{katz2022sra}. Its records link studies, biological samples, experiments and sequencing runs. The experiment metadata include a free-text \texttt{library\_construction\_protocol} field intended to describe the preparation of sequencing libraries \cite{ncbiSraMetadata}. Whether the information in this and related fields is sufficiently detailed for later analysis depends on what submitters record.

Standards such as MIAME and MINSEQE specify information to be reported about experiments. MAGE-ML and MAGE-TAB, by comparison, provide formats for representing and exchanging experimental metadata \cite{brazma2009miame,rayner2006magetab}. Experience with MIAME showed that the usability of a reporting format can influence its adoption \cite{brazma2009miame}. A structured submission system therefore does not, by itself, ensure that the protocol descriptions are complete.

The 2015 investigation by Alnasir and Shanahan examined the annotation of fragmentation, adapter ligation and enrichment using the Bioconductor SRAdb SQLite metadata database extracted on 3 December 2013 \cite{alnasir2015annotation,zhu2013sradb}. The original analysis searched relevant free-text fields using three manually assembled lists of related terms. In the \texttt{library\_construction\_protocol} field, 7.10\% of experiment records matched the fragmentation list, 5.84\% matched adapter ligation and 7.57\% matched enrichment. Only 4.06\% matched all three groups. Approximately half the records had a NULL value in that field. The study-level analysis also showed that protocol-related terms could be present in only some of the experiments belonging to a study \cite{alnasir2015annotation}. These measurements quantified the occurrence of selected terms; they did not assess whether a protocol description contained sufficient information to reproduce the laboratory procedure.

Stevens et al. subsequently cited these results in their recommendations for annotating sequencing experiments, identifying fragmentation, adapter ligation and enrichment as details that should be considered during experimental planning \cite{stevens2020rules}.

The SRA has since grown substantially, and library preparation methods have changed \cite{katz2022sra}. Transposase-mediated preparation and amplification-free protocols, for example, have altered both the steps used and the terminology by which they are described. BioSample and BioProject now provide additional linked metadata resources \cite{ncbiBioSample}. The extent to which these developments have improved the annotation of library preparation in experiment records remains uncertain.

Studies of SRA-associated metadata continue to identify problems with missing information and inconsistent terminology. Klie et al. examined heterogeneity in BioSample attributes and investigated the recovery of information from existing text \cite{klie2021metadata}. The FAIR principles similarly emphasise the need for metadata that can support reuse and interoperability \cite{wilkinson2016fair}. For sequencing protocols, however, the presence of descriptive text is insufficient if the methods cannot be identified or compared reliably.

The original investigation also provides an opportunity to examine long-term computational reproducibility. Concerns about the reproducibility of sequencing analysis were already established when it was published \cite{nekrutenko2012reproducibility}. The original authors deposited the SQL scripts and identified the metadata snapshot used to generate the results \cite{alnasir2015annotation,alnasir2015software}. Executing those scripts on the original database would test reproduction of the historical findings. Applying them to newly acquired metadata asks a different question: whether the same analytical definitions remain usable after substantial changes in the archive. Even with identical table names, differences in record populations and field mappings can affect the comparison.

This study applies the published SQL keyword expressions, unchanged, to a September 2026 NCBI metadata snapshot reconstructed in the original SRAdb relational schema. It compares the frequency of matches in the fields examined in 2015 and assesses their distribution among studies and experiments. The original term lists provide a fixed basis for comparison, although their sensitivity to newer library preparation methods is likely to differ. Expanded terminology is therefore kept separate from the historical analysis. The study also stratifies contemporary experiment-level matches by sequencing library strategy and examines whether the surviving metadata and SQL artefacts support the original methods of analysis. The aim is to establish how the detectability and location of the three historical protocol categories have changed, and to identify the methodological constraints on interpreting those differences.
\section{Materials and Methods}

\subsection{Historical dataset and published SQL}
The historical reference comprised the SRAdb SQLite snapshot extracted on
3 December 2013 and described in Alnasir and Shanahan (2015)
\cite{alnasir2015annotation,zhu2013sradb}. The published SQL archive \cite{alnasir2015software} contains the
keyword lists and queries used for the original field-level and
study-level analyses, together with supplementary queries. Historical
counts were taken from the published article and, where explicitly
identified, from the deposited SQL comments. The 2015 keyword predicates were not altered when generating the
contemporary results. The queries have not been independently rerun
against the original database. The contemporary results therefore demonstrate reuse of
the historical analytical definitions, while independent reproduction
of the published 2015 counts remains outstanding.

The original published lists are reproduced in Table~\ref{tab:historical-keyword-lists}.
They are shown as printed in the 2015 article; the deposited executable
SQL predicates remain the specification used for the contemporary analyses.

\begin{table}[htbp]
\centering
\caption{Original structured keyword lists for the three sequencing
protocol-step groups, reproduced from Table~1 of Alnasir and Shanahan
(2015) \cite{alnasir2015annotation}. The percent sign (\%) denotes
a wildcard in an SQL \texttt{LIKE} expression. The terms are
transcribed as published; the archived executable SQL takes precedence
where its predicates differ from the printed table.}
\label{tab:historical-keyword-lists}
\small
\begin{tabular}{@{}lll@{}}
\toprule
\textbf{Fragmentation} & \textbf{Adapter ligation} & \textbf{Enrichment} \\
\midrule
\texttt{shear}         & \texttt{adapter}           & \texttt{clone\%} \\
\texttt{restriction}   & \texttt{ligat\%}          & \texttt{clonin\%} \\
\texttt{digest}        & \texttt{blunt\%}          & \texttt{vector\%} \\
\texttt{fragment}      & \texttt{phosphorylat\%}   & \texttt{pcr} \\
\texttt{breaks}        & \texttt{overhang}          & \texttt{amplif\%} \\
\texttt{acoustic}      & \texttt{t4-pnk}            & \texttt{polymerase} \\
\texttt{nebulisation}  & \texttt{t4}                & \texttt{taq} \\
\texttt{nebulization}  & \texttt{pnk}               & \texttt{phusion} \\
\texttt{nebuliz}       & \texttt{kinase}            & \texttt{temperat\%} \\
\texttt{nebulis}       & \texttt{a-tail}            & \texttt{thermal\%} \\
\texttt{sonic}         &                         & \texttt{anneal\%} \\
                      &                         & \texttt{denature\%} \\
\bottomrule
\end{tabular}
\end{table}

\subsection{Metadata acquisition, provenance and relational database reconstruction}
\label{subsec:metadata_provenance}
SRA metadata were obtained directly from the National Center for
Biotechnology Information (NCBI) full metadata archive, which was
retained as the fixed source dataset for the contemporary analysis.
The archive contained 21,440,783 XML files representing approximately
218.6 GiB of uncompressed XML metadata. The archive provenance and
processing environment are summarised in Table~\ref{tab:sra-provenance}.

\begin{table}[htbp]
\centering
\caption{Provenance and processing environment for the September 2026
NCBI SRA metadata snapshot.}
\label{tab:sra-provenance}
\small
\begin{tabularx}{\linewidth}{@{}p{0.24\linewidth}X@{}}
\toprule
\textbf{Property} & \textbf{Recorded value} \\
\midrule
Source archive & \texttt{NCBI\_SRA\_Metadata\_Full\_20260913.tar.gz} \\
Snapshot date & 13 September 2026 \\
Archive size & 18,233,005,318 bytes (16.98 GiB) \\
SHA-256 & {\footnotesize\ttfamily 827a730774db5435e0a75aecccd458f741f535cf33aa9fda6553367b3cf5433b} \\
DB type & SQLite3 \\
Processing platform & Three-node Ubuntu Linux/OpenLava cluster \\
\bottomrule
\end{tabularx}
\end{table}

Custom Python scripts mapped NCBI XML metadata for submissions,
studies, samples, experiments and sequencing runs to a SQLite3
database reproducing the original Bioconductor SRAdb table
and field names. Processing used a 20~Gbit/s InfiniBand
interconnect and the OpenLava batch-job scheduler. The XML input
was partitioned into three accession-based shards, each generating
a local SQLite database; the databases were subsequently consolidated
by accession identifiers. The reported record counts are those
obtained from the consolidated database. Field-level mapping
and accession relationships were evaluated as described below;
the source-level submission-date comparison was examined separately
from validation of the library preparation fields.

The reconstruction retained the SRAdb table and column definitions.
The auxiliary \texttt{sra} table was unpopulated, however, and several
SRAdb-specific fields, including archive links and
\texttt{sradb\_updated}, were NULL throughout. None of these fields
was needed for the principal annotation queries. The reconstructed
database is structurally compatible with SRAdb, although its derived
metadata have not been reproduced in full.

\subsection{Independent validation of reconstructed experiment fields}
\label{subsec:experiment-xml-validation}
The reconstruction of
\texttt{experiment.design\_description},
\texttt{experiment.library\_construction\_protocol} and
\texttt{experiment.library\_strategy} was independently checked against the source
\texttt{EXPERIMENT} XML. A deterministic purposive sample of
150 accessions was drawn from six mutually exclusive
experiment-level keyword-classification groups, including
rare both-field and cross-field-only matches. Selection also
covered multiple library strategies and accession series.
The two description fields and library strategy were
extracted independently from the original XML, compared
with their SQLite values and evaluated using the historical
keyword predicates.

All 150 accessions were located, and every XML-derived
historical keyword classification agreed with SQLite. Across the
three fields, 100 records agreed exactly; 25 exhibited
whitespace-only differences, and in 25 the XML extractor returned zero-length
design-description text where SQLite stored NULL. No substantive
text-content or library-strategy discrepancies were identified
in the sample. The strict text-equality audit therefore remained
flagged for review, notwithstanding complete keyword-classification
agreement. This purposive audit supports the reconstruction of the
fields used in the selected comparisons; it cannot estimate
archive-wide mapping error or establish whether keyword matches
represent informative laboratory protocols. Sampling definitions,
category-level results and strict audit status are documented in
the Supplementary Information (``Independent source-XML
validation of experiment fields'').

\subsection{Original vocabulary and counting definitions}
The original keyword lists for fragmentation, adapter ligation
and enrichment (Table~\ref{tab:historical-keyword-lists}) were
compiled through manual review of sequencing library preparation
methods \cite{alnasir2015annotation}. The analysis retained
the deposited \texttt{LIKE} predicates, including their wildcard
placements and original Boolean expressions. A match records the presence of a search term. It provides no
direct assessment of the protocol's experimental detail or its
applicability to the sequencing method.
The original investigation did not estimate the false-negative
rate of keyword detection.

To complete the September 2026 equivalent of the original Table~3,
a separately identified supplementary query was used,
\texttt{SQL-SRA-QRY-wordlists-all-fields-three-step-2026.txt},
for the five fields lacking all-three-step counts in the supplied
unjoined output. The query combined the deposited field-specific
keyword predicates without changing their search expressions.
The all-three-step count in \texttt{library\_construction\_protocol}
came from the original archived query; the value for
\texttt{library\_selection} is zero because no ligation or
enrichment matches were detected in that field. The archived
fragmentation predicate for \texttt{experiment\_attribute} refers
once to \texttt{library\_selection} (for \texttt{nebuliz}).
That reference was retained in the supplementary historical-equivalent
calculation, so its resulting intersection is not strictly confined
to \texttt{experiment\_attribute}.

Field-level percentages use the total number of rows in the
appropriate \texttt{study}, \texttt{sample} or \texttt{experiment}
table. The all-three-step experiment metric requires all three
keyword groups in the same \texttt{library\_construction\_protocol}
value. For study-level measures, the query first joins studies
to experiments on \texttt{study\_accession}, then groups by study.
For the study-level all-three-step count, at least one experiment
must match all three groups. Matches distributed across several
experiments are excluded from that count. A study is classified as unannotated when none of its joined
experiments has a match in the library-construction field. This
definition does not include possible protocol information in other
metadata fields.

\subsection{Supplementary analyses and integrity checks}
The deposited study-level pairwise co-occurrence,
within-study annotation distribution, NULL-field, URL, cross-field
and submission-year SQL analyses were executed against the contemporary
database. The distribution analysis bins positive studies by the
fraction of their experiments containing matches into
$(0,25]$, $(25,50]$, $(50,75]$ and $(75,100]$ percent intervals.
Those proportions are conditional upon at least one matching
experiment. The pairwise queries identify studies containing an experiment
that matches both keyword groups. Their counts differ from
conventional study-level intersections when the terms appear
in separate experiments.

Additional SQLite queries counted major relational records,
examined missing accession links and evaluated submission-date
availability by accession series. The deposited historical
cross-field query mixes \texttt{OR} and \texttt{AND}
expressions in a way that prevents its counts from
representing experiment-level overlap. To obtain valid experiment-level overlap counts, a separate
supplementary query was used:
\texttt{SQL-SRA-QRY-experiment-cross-field-complementarity-2026.txt}.
It evaluates the original field-specific keyword
predicates independently for
\texttt{library\_construction\_protocol} and
\texttt{design\_description} within each experiment
record. For each keyword group, an experiment was
classified as matching the protocol field only, the
design description only, both or neither. The
all-three-step field-specific comparison requires all three
keyword groups within an individual field. The analysis also
tested whether the three groups could be detected across
the two fields collectively, even when neither field
contained all three groups independently.
Design-only matches were subdivided according to whether
the protocol field was NULL. For the collective all-three-step
criterion, let $N_{C}$ be the number of experiments with all
three groups detectable across the two fields and $N_{P}$
the number with all three in the protocol field alone.
The proportion missed by a protocol-field-only search was
calculated as $(N_{C}-N_{P})/N_{C}$. This denominator is
restricted to experiments meeting the historical criterion
across the two fields; the measure is not recall against
independently established, scientifically adequate protocols.
All marginal counts were reconciled with the earlier
field-specific SQL outputs and pre-flight database counts.

\subsection{Source-level submission-date check}
The presence of \texttt{submission\_date} in all 7,891,372
source \texttt{SUBMISSION} elements was checked independently
against the corresponding SQLite values. The INSDC SRA submission
XML specification (version~1.5.9) was used to interpret this
optional, submitter-assigned date
\cite{insdcSraSubmissionSchema}; it was not treated as an
archive receipt timestamp. Further accession-level and
year-distribution details are provided in the
Supplementary Information.

\subsection{Stratification by sequencing library strategy}
Experiment-level keyword classifications were stratified by
\texttt{experiment.library\_strategy} using the supplementary query
\texttt{SQL-SRA-QRY-experiment-library-strategy-stratification-2026.txt}.
The query retained the original field-specific predicates and
calculated, within each recorded strategy, the numbers of
experiments matching all three keyword groups in
\texttt{library\_construction\_protocol}, in
\texttt{design\_description}, and across the two fields collectively.
NULL values in the protocol field were also counted. For each strategy,
the denominator was its total number of experiment records; the
collective criterion permitted the three groups to be distributed
between the fields. No equivalence between a keyword match and an
applicable or sufficiently described laboratory procedure was assumed.

Strategies with at least 100,000 experiments were reported individually
in the main table; less frequent named strategies were pooled for
presentation only, and their individual counts were retained in the
full supplementary output. The SRA category \texttt{OTHER} was kept
distinct from this pooled group. The stratified totals were checked against the experiment-level
cross-field analysis.

\subsection{Comparative interpretation}
The published 2013 baseline and the 2026 measurements are reported
as counts and proportions. The vocabulary is held constant
across snapshots, although the underlying populations differ
substantially in size and sequencing assay composition. The results are interpreted as changes in \emph{historical keyword-detected coverage}.
The contemporary strategy-level analysis characterises heterogeneity
within the 2026 snapshot; without equivalent historical strata, it
cannot apportion the observed decade-long change to shifts in strategy
composition. Significance tests were not applied to the archive-wide
counts. The presence of a term alone cannot establish the adequacy
of an experimental protocol.

\section{Results}

\subsection{Growth of the archive and population of protocol fields}
The reconstructed September 2026 database contained 756,894 studies,
43,510,050 samples, 41,593,302 experiments, 44,224,992 runs and
7,891,372 submissions. The contemporary experiment population was
99.1 times that reported for December 2013 (419,620), while the
study population was 25.6 times larger (29,598 in 2013)
\cite{alnasir2015annotation}.

Of the 41,593,302 contemporary experiments, 28,138,529 (67.65\%)
had a NULL \texttt{library\_construction\_protocol} field, leaving
13,454,773 (32.35\%) non-NULL values. There were no exact
empty-string values in this field (whitespace-only strings were not separately counted). The 2015 paper reported
212,070 NULL fields (51.12\%), using a denominator of 414,788
experiments in its narrative, rather than the 419,620 in
Table~3 (which gives 50.54\% for the same NULL count).
The contemporary NULL proportion exceeds either historical
figure. The \texttt{design\_description} field was NULL in
13,947,757 contemporary experiments (33.53\%).

\subsection{Change in historical keyword coverage}
\label{subsec:result-keywords}
In the designated \texttt{library\_construction\_protocol}
field, the original fragmentation, adapter-ligation and
enrichment predicates detected matches in 1,683,325
(4.05\%), 918,834 (2.21\%) and 2,502,137 (6.02\%)
contemporary experiments, respectively. All three groups
matched in the same field for 591,526 experiments (1.42\%).
The published 2013 percentages were 7.10\%, 5.84\%,
7.57\% and 4.06\% (Table~\ref{tab:protocol-coverage}).
All-three-step keyword coverage fell by 2.63 percentage
points, corresponding to a 64.9\% relative reduction.
The absolute count increased 34.8-fold over the same period,
compared with 99.1-fold growth in the experiment population.

\begin{table}[htbp]
\centering
\caption{Historical keyword matches in
\texttt{experiment.library\_construction\_protocol}.
The 2013 counts are from Table 3 of Alnasir and Shanahan
(2015); 2026 counts are from the supplied SQL output.
Percentages use 419,620 and 41,593,302 experiments,
respectively.}
\label{tab:protocol-coverage}
\normalsize
\begin{tabular}{@{}lrrrr@{}}
\toprule
& \multicolumn{2}{c}{2013 (published)} &
  \multicolumn{2}{c}{2026 (queried)}\\
\cmidrule(lr){2-3}\cmidrule(l){4-5}
Keyword group & Count & \% & Count & \%\\
\midrule
Fragmentation & 29,799 & 7.10 & 1,683,325 & 4.05\\
Adapter ligation & 24,486 & 5.84 & 918,834 & 2.21\\
Enrichment & 31,782 & 7.57 & 2,502,137 & 6.02\\
All three in one field & 17,021 & 4.06 & 591,526 & 1.42\\
\bottomrule
\end{tabular}
\end{table}

\begin{figure}[htbp]
\centering
\includegraphics[width=\linewidth]{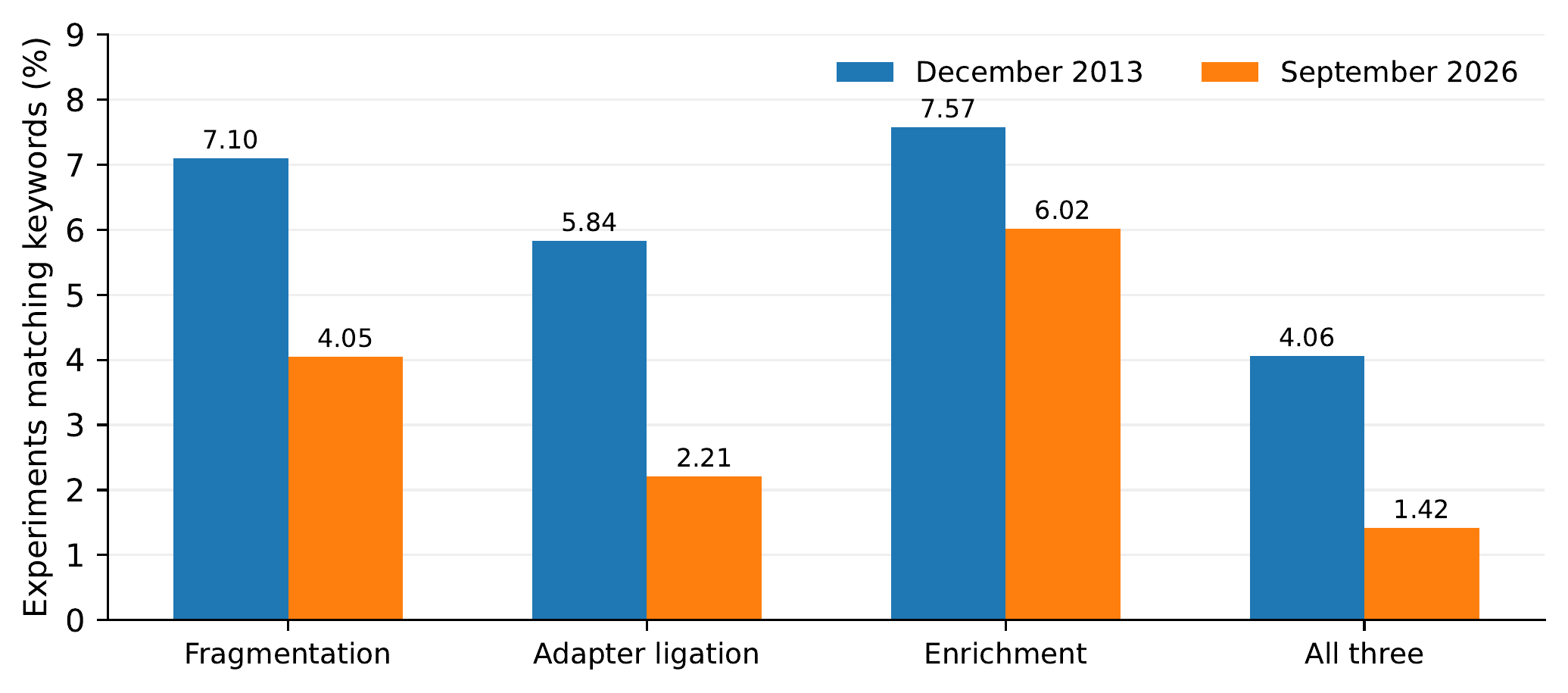}
\caption{Percentage of experiment records containing historical
protocol-step keywords in the designated library construction
field. The two snapshots were evaluated using the deposited
2015 keyword definitions; the percentages do not assess the
adequacy of the underlying protocol descriptions.}
\label{fig:protocol-coverage}
\end{figure}

\subsection{Changes in the distribution of matches across metadata fields}
Keyword coverage differed markedly between metadata fields
(Table~\ref{tab:field-coverage}). In
\texttt{experiment.design\_description}, fragmentation
coverage increased from 2.79\% to 4.64\%, adapter
ligation from 1.52\% to 3.66\%, and enrichment from
4.00\% to 9.79\%. Each of these percentages exceeded
its contemporary counterpart in
\texttt{library\_construction\_protocol}. The all-three-step
counts across the same fields are reported below.
Study-abstract
coverage also increased for the three categories, whereas
coverage in \texttt{study\_description} and
\texttt{experiment\_attribute} decreased. The direction of change therefore depends on which
metadata field is examined.

\begin{table}[htbp]
\centering
\caption{Historical keyword coverage by metadata field.
Values are the percentages of records matching each group
in 2013~$\rightarrow$~2026. Denominators are the totals for
the relevant study, sample or experiment table in each
snapshot. Historical rates are recalculated from the
published counts in Table~3.}
\label{tab:field-coverage}
\footnotesize
\setlength{\tabcolsep}{5pt}
\begin{tabular}{@{}lccc@{}}
\toprule
Field & Fragmentation & Adapter ligation & Enrichment\\
\midrule
\texttt{study\_abstract} & 1.27 $\rightarrow$ 2.24 &
0.47 $\rightarrow$ 1.79 & 3.18 $\rightarrow$ 4.56\\
\texttt{study\_description} & 0.99 $\rightarrow$ 0.33 &
0.46 $\rightarrow$ 0.11 & 1.65 $\rightarrow$ 0.72\\
\texttt{sample.description} & 0.34 $\rightarrow$ 0.45 &
0.19 $\rightarrow$ 0.20 & 0.45 $\rightarrow$ 0.82\\
\texttt{design\_description} & 2.79 $\rightarrow$ 4.64 &
1.52 $\rightarrow$ 3.66 & 4.00 $\rightarrow$ 9.79\\
\texttt{library\_selection} & 0.36 $\rightarrow$ 1.73 &
0.00 $\rightarrow$ 0.00 & 0.00 $\rightarrow$ 0.00\\
\texttt{library\_construction\_protocol} & 7.10 $\rightarrow$ 4.05 &
5.84 $\rightarrow$ 2.21 & 7.57 $\rightarrow$ 6.02\\
\texttt{experiment\_attribute} & 0.10 $\rightarrow$ 0.02 &
0.24 $\rightarrow$ 0.08 & 0.67 $\rightarrow$ 0.12\\
\bottomrule
\end{tabular}
\end{table}

The combined three-step criterion also differed between fields.
In the 2026 experiment records, 606,104 (1.46\%) matched all
three historical keyword groups in \texttt{design\_description},
compared with 591,526 (1.42\%) in
\texttt{library\_construction\_protocol}. The corresponding
published 2013 counts were 2,691 (0.64\%) and 17,021
(4.06\%), respectively \cite{alnasir2015annotation}.
The complete contemporary field-level counts are given in
Table~\ref{tab:field-counts-2026}. Their overlap within
individual experiments is examined below.

\begin{table}[htbp]
\centering
\caption{Historical protocol-step keyword matches across
the September 2026 SRA metadata fields, corresponding to
Table~3 of Alnasir and Shanahan (2015). Counts are shown with
percentages of the respective table populations
(study: 756,894; sample: 43,510,050; experiment: 41,593,302).
The last column requires all three keyword groups, using the
deposited predicates; see Methods for the additional SQL
and the anomalous \texttt{experiment\_attribute} predicate.}
\label{tab:field-counts-2026}
\footnotesize
\setlength{\tabcolsep}{3pt}
\renewcommand{\arraystretch}{1.13}
\begin{tabular}{@{}llcccc@{}}
\toprule
Table & Field & Fragmentation & Adapter ligation & Enrichment & All three\\
\midrule
Study & \texttt{study\_abstract} &
\shortstack{16,921\\(2.24\%)} &
\shortstack{13,563\\(1.79\%)} &
\shortstack{34,520\\(4.56\%)} &
\shortstack{786\\(0.10\%)} \\
Study & \texttt{study\_description} &
\shortstack{2,526\\(0.33\%)} &
\shortstack{815\\(0.11\%)} &
\shortstack{5,458\\(0.72\%)} &
\shortstack{126\\(0.02\%)} \\
Sample & \texttt{description} &
\shortstack{197,401\\(0.45\%)} &
\shortstack{85,097\\(0.20\%)} &
\shortstack{357,160\\(0.82\%)} &
\shortstack{54,122\\(0.12\%)} \\
Experiment & \texttt{design\_description} &
\shortstack{1,931,685\\(4.64\%)} &
\shortstack{1,523,161\\(3.66\%)} &
\shortstack{4,072,157\\(9.79\%)} &
\shortstack{606,104\\(1.46\%)} \\
Experiment & \texttt{library\_selection} &
\shortstack{719,783\\(1.73\%)} &
\shortstack{0\\(0\%)} &
\shortstack{0\\(0\%)} &
\shortstack{0\\(0\%)} \\
Experiment & \shortstack[l]{\texttt{library\_construction\_}\\\texttt{protocol}} &
\shortstack{1,683,325\\(4.05\%)} &
\shortstack{918,834\\(2.21\%)} &
\shortstack{2,502,137\\(6.02\%)} &
\shortstack{591,526\\(1.42\%)} \\
Experiment & \texttt{experiment\_attribute} &
\shortstack{7,648\\(0.02\%)} &
\shortstack{31,366\\(0.08\%)} &
\shortstack{50,009\\(0.12\%)} &
\shortstack{369\\(0.0009\%)} \\
\bottomrule
\end{tabular}
\end{table}

\subsection{Cross-field complementarity of protocol-step keyword matches}
\label{subsec:cross-field-complementarity}
Historical keyword matches were compared between
\texttt{library\_construction\_protocol} and
\texttt{design\_description} within each experiment
record (Table~\ref{tab:cross-field-complementarity}).
Most matching experiments were detected in only one of
the two fields. For the combined three-step criterion,
591,179 experiments matched exclusively in the protocol
field, 605,757 exclusively in the design description and
347 in both fields.

\begin{table}[htbp]
\centering
\caption{Overlap of historical protocol-step keyword matches
between \texttt{library\_construction\_protocol} (P)
and \texttt{design\_description} (D), September 2026.
The four categories are mutually exclusive and each row
accounts for all 41,593,302 experiment records. For the
last row, each field must independently contain matches
to all three historical keyword groups. Thus, the
``Neither'' category in this row includes the 84 experiments
whose three groups can be detected only by combining
partial matches across the two fields.}
\label{tab:cross-field-complementarity}
\footnotesize
\setlength{\tabcolsep}{6pt}
\renewcommand{\arraystretch}{1.15}
\begin{tabular}{@{}lrrrr@{}}
\toprule
Keyword group & P only & Both & D only & Neither\\
\midrule
Fragmentation & 1,682,059 & 1,266 & 1,930,419 & 37,979,558\\
Adapter ligation & 909,024 & 9,810 & 1,513,351 & 39,161,117\\
Enrichment & 2,482,961 & 19,176 & 4,052,981 & 35,038,184\\
All three steps & 591,179 & 347 & 605,757 & 40,396,019\\
\bottomrule
\end{tabular}
\end{table}

Searching the protocol field alone identified 591,526
experiments (1.42\%) satisfying the all-three-step
criterion. Searching both fields collectively identified
1,197,367 (2.88\%). Consequently, a protocol-field-only
search detected 49.40\% of experiments satisfying the
historical criterion across these two fields and missed
605,841 (50.60\%). The additional matches represent
a 102.42\% increase relative to the protocol-field-only
count, or 1.46 percentage points of the entire experiment
population. Of the collective matches, 1,197,283 contained
all three groups within at least one field independently;
only 84 required partial matches from both fields.
The 50.60\% figure is a conditional missed-detection rate
for the historical keywords, not an estimate of missing
adequate laboratory protocols.

The extra matches were concentrated among experiments
without a populated protocol field. Of the 605,757
experiments matching all three groups exclusively in
\texttt{design\_description}, 604,116 (99.73\%) had
a NULL \texttt{library\_construction\_protocol} value.
The other 1,641 had a populated protocol field that did
not meet the combined three-step criterion. All 11
internal reconciliation checks against the established
2026 totals passed.
The partition of all-three-step detections is illustrated in
Figure~\ref{fig:sra-field-detection-panels}A.

\subsection{Variation by sequencing library strategy}
\label{subsec:strategy-coverage}
The contemporary three-step matching rate varied substantially by
sequencing library strategy (Table~\ref{tab:strategy-coverage}).
The three most common categories, \texttt{AMPLICON}, \texttt{WGS}
and \texttt{RNA-Seq}, accounted for 82.33\% of experiment records.
Their collective two-field matching rates were 0.26\%, 2.29\%
and 7.69\%, respectively. Although \texttt{AMPLICON} accounted for
43.62\% of experiments, it contributed 46,828 (3.91\%) of
the 1,197,367 collective matches. By comparison,
\texttt{RNA-Seq} contributed 540,985 (45.18\%) of these matches
while representing 16.92\% of experiment records.

\begin{figure}[p]
\centering
\includegraphics[
    width=0.98\linewidth,
    height=0.28\textheight,
    keepaspectratio
]{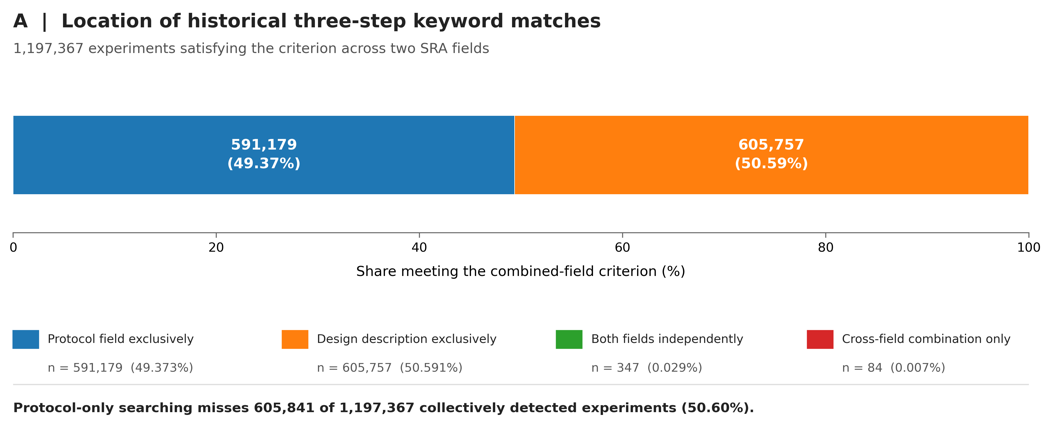}

\vspace{0.5em}

\includegraphics[
    width=0.92\linewidth,
    height=0.48\textheight,
    keepaspectratio
]{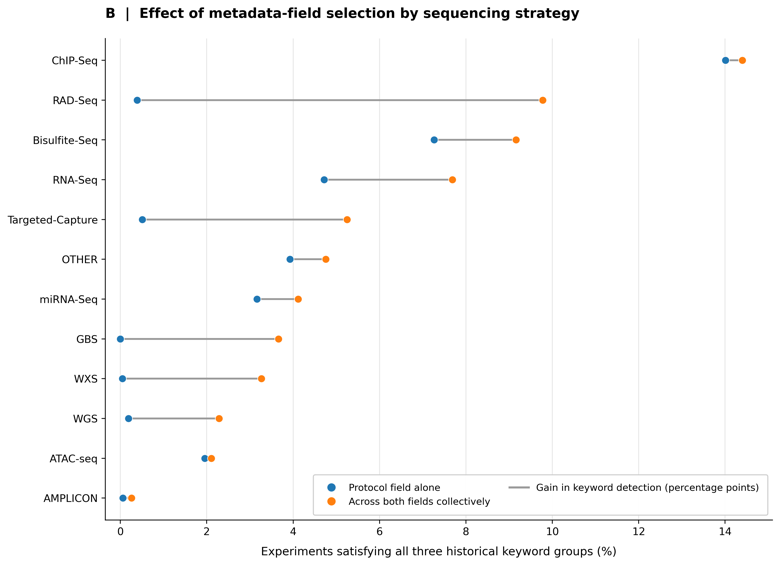}
\caption{Distribution of historical three-step keyword matches between
SRA experiment metadata fields, September 2026.
\textbf{(A)} Among 1,197,367 experiments in which fragmentation,
adapter ligation and enrichment keywords were collectively detectable
across \texttt{library\_construction\_protocol} and
\texttt{design\_description}, 591,179 matched all three groups
exclusively in the protocol field, 605,757 exclusively in the design
description and 347 independently in both fields. A further 84
required complementary matches across the two fields.
The green (347; 0.029\%) and red (84; 0.007\%) segments are too
small to be visually resolved in the proportional stacked bar;
their values are shown in the figure key.
A protocol-field-only search misses 605,841 of these
experiments (50.60\%).
\textbf{(B)} For each displayed sequencing library strategy, blue
points indicate the percentage of experiments meeting the
three-step criterion in the protocol field alone; orange points
indicate the percentage meeting it collectively across both fields.
Grey connecting lines show the increase in percentage points
from searching both fields. Percentages in panel B use the total
number of experiments within each strategy as their denominators.
Both panels measure detection of the historical keyword groups,
not the completeness or scientific adequacy of laboratory protocols.}
\label{fig:sra-field-detection-panels}
\end{figure}

The relative contribution of the two text fields also differed
by strategy. For \texttt{ChIP-Seq}, 14.01\% of experiments
matched all three groups in the designated protocol field and
0.39\% in the design description. In \texttt{RAD-Seq}, the
corresponding rates were 0.39\% and 9.39\%. The proportion
of NULL protocol fields was 18.16\% for \texttt{ChIP-Seq}
and 95.18\% for \texttt{RAD-Seq}. These differences show
that the archive-wide rates conceal considerable variation
in both keyword detection and field population across
library strategies.
The strategy-level differences are shown in
Figure~\ref{fig:sra-field-detection-panels}B.

\begin{table}[htbp]
\centering
\caption{Historical all-three-step keyword detection by
sequencing library strategy, September 2026. P and D give
the percentages of experiments matching all three groups
within \texttt{library\_construction\_protocol} and
\texttt{design\_description}, respectively. Combined
permits the three groups to occur anywhere across the two fields;
NULL P is the percentage with a missing protocol field.
Percentages use each row's experiment count as denominator.
\texttt{OTHER} is an original SRA strategy value, distinct
from the pooled row of smaller named strategies.}
\label{tab:strategy-coverage}
\footnotesize
\setlength{\tabcolsep}{4pt}
\renewcommand{\arraystretch}{1.1}
\begin{tabular}{@{}lrrrrr@{}}
\toprule
Library strategy & Experiments & P (\%) & D (\%) & Combined (\%) & NULL P (\%)\\
\midrule
\texttt{AMPLICON} & 18,141,462 & 0.06\% & 0.20\% & 0.26\% & 79.81\%\\
\texttt{WGS} & 9,065,085 & 0.19\% & 2.10\% & 2.29\% & 70.96\%\\
\texttt{RNA-Seq} & 7,037,816 & 4.72\% & 2.97\% & 7.69\% & 42.27\%\\
\texttt{OTHER} & 2,804,190 & 3.93\% & 0.83\% & 4.76\% & 32.38\%\\
\texttt{RAD-Seq} & 768,937 & 0.39\% & 9.39\% & 9.78\% & 95.18\%\\
\texttt{Targeted-Capture} & 639,356 & 0.51\% & 4.74\% & 5.25\% & 90.27\%\\
\texttt{WXS} & 553,857 & 0.05\% & 3.22\% & 3.27\% & 88.92\%\\
\texttt{WGA} & 450,922 & 0.02\% & 1.09\% & 1.10\% & 96.22\%\\
\texttt{ChIP-Seq} & 408,532 & 14.01\% & 0.39\% & 14.40\% & 18.16\%\\
\texttt{Bisulfite-Seq} & 293,919 & 7.27\% & 1.89\% & 9.16\% & 57.56\%\\
\texttt{CLONE} & 288,790 & 0.00\% & 0.02\% & 0.02\% & 73.14\%\\
\texttt{ATAC-seq} & 218,783 & 1.95\% & 0.15\% & 2.10\% & 25.30\%\\
\texttt{miRNA-Seq} & 189,829 & 3.16\% & 0.95\% & 4.12\% & 46.24\%\\
\texttt{GBS} & 178,936 & 0.00\% & 3.66\% & 3.66\% & 94.14\%\\
Other named ($<100{,}000$ each) & 552,888 & 4.81\% & 1.05\% & 5.86\% & 62.11\%\\
\midrule
\textbf{All experiments} & \textbf{41,593,302} & \textbf{1.42\%} & \textbf{1.46\%} & \textbf{2.88\%} & \textbf{67.65\%}\\
\bottomrule
\end{tabular}
\end{table}

\subsection{Study-level annotation and co-occurrence}
\label{subsec:result-studies}
The study--experiment join identified 711,671 distinct
study accession groups in the contemporary database.
At least one experiment matched the historical
fragmentation, adapter-ligation and enrichment predicates
in 56,759 (7.98\%), 32,322 (4.54\%) and
56,526 (7.94\%) studies, respectively. For all three
groups within one experiment, the corresponding result
was 22,096 studies (3.10\% of joined studies;
2.92\% of all 756,894 study records). The 2015
paper reported 1,409 all-three-step studies
(4.76\% of its 29,598 studies).

Among the contemporary joined studies, 631,299
(88.71\%) had no match to any of the three groups
in the designated protocol field. The original paper
reported 21,799 of 29,598 studies (73.65\%)
without detectable annotation. The historical
denominator comprises all study records, whereas
the contemporary denominator comprises studies
joined to experiments; the scope of the historical
absence criterion also requires care when making
a direct numerical comparison.

The pairwise contemporary queries identified 25,088
studies with fragmentation--ligation matches, 35,521
with fragmentation--enrichment and 26,524 with
ligation--enrichment. Their respective percentages
of joined studies were 3.53\%, 4.99\% and 3.73\%.
The original repository comments reported 1,580,
1,857 and 1,709 for the same three combinations
in the historical snapshot. These counts require pairwise matches within a single
experiment. They cannot be used as exclusive regions
of a study-level Venn diagram, since different pairs
may occur in different experiments within the same study.

\subsection{Within-study distribution of annotation}
\label{subsec:within-study}
Applying the historical within-study distribution analysis
to the September 2026 dataset identified 22,096 studies
with at least one experiment matching all three
historical keyword groups. Of these, 21,058
(95.30\%) had matches in more than 75\% of
their experiments (Table~\ref{tab:within-study}).
Comparable high conditional proportions were
observed for fragmentation (96.23\%), ligation
(95.41\%) and enrichment (96.44\%). By design,
studies containing a single matching experiment
are classified as having 100\% within-study
coverage when that experiment is the only one
associated with the study. The four bins show a concentration of matches among
keyword-positive studies. The influence of study size on
this distribution has not yet been quantified.

\begin{table}[htbp]
\centering
\caption{Number of contemporary studies by percentage
of their experiments matching the historical keyword
groups. Only studies containing at least one matching
experiment are included; percentages refer to
experiments within a study. The final column is the
fraction of positive studies in the $>75$--100\% bin.}
\label{tab:within-study}
\normalsize
\begin{tabular}{@{}lrrrrr@{}}
\toprule
Group & $(0,25]$ & $(25,50]$ & $(50,75]$ & $(75,100]$ & High (\%)\\
\midrule
Fragmentation & 590 & 854 & 695 & 54,620 & 96.23\\
Adapter ligation & 509 & 561 & 414 & 30,838 & 95.41\\
Enrichment & 601 & 783 & 628 & 54,514 & 96.44\\
All three & 356 & 394 & 288 & 21,058 & 95.30\\
\bottomrule
\end{tabular}
\end{table}

\begin{figure}[htbp]
    \centering
    \includegraphics[width=\linewidth,
                     height=0.65\textheight,
                     keepaspectratio]{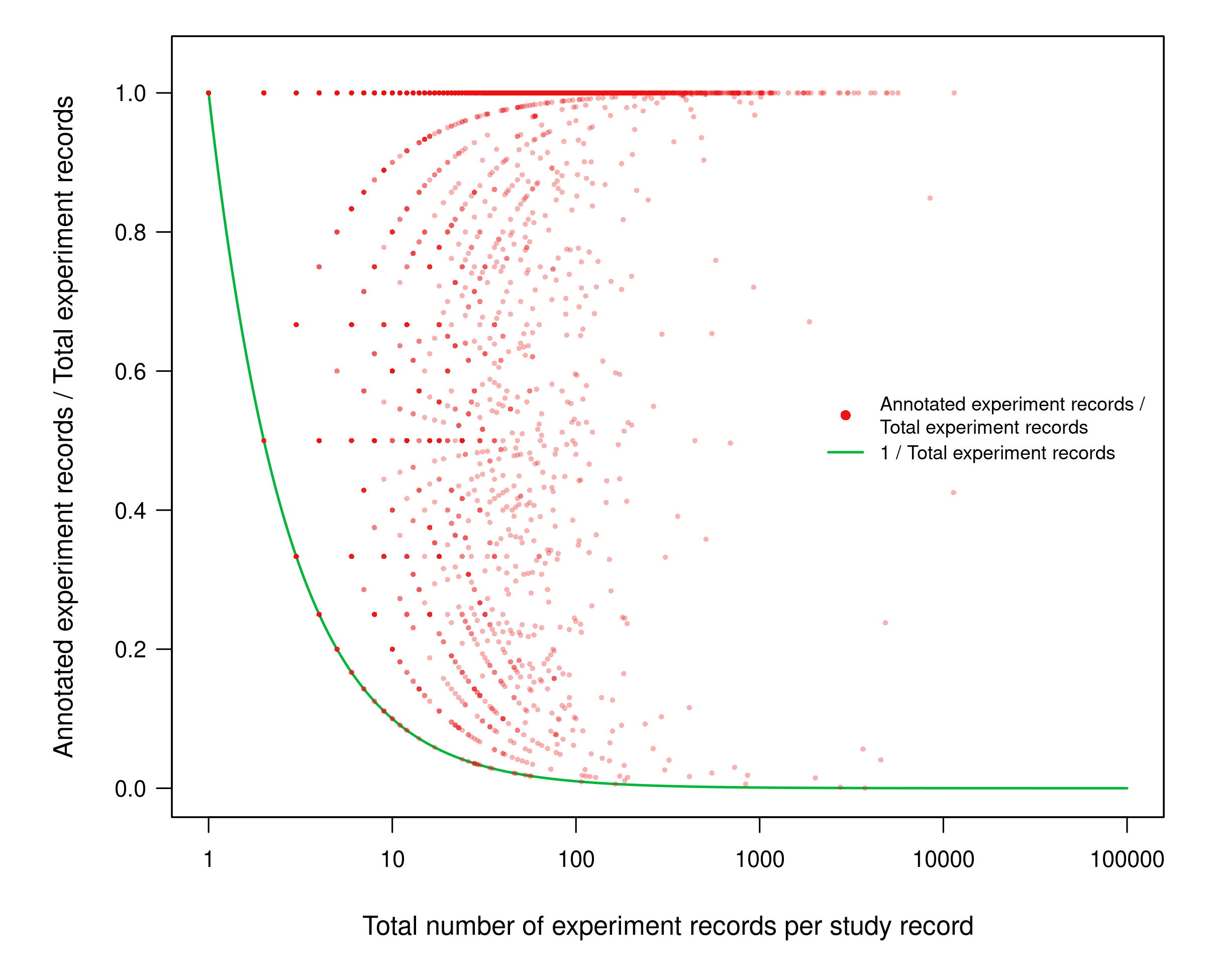}
    \caption{Proportion of experiments matching all three
    historical protocol-step keyword groups within individual SRA
    studies, September 2026. For each study, the fraction of
    matching experiments is plotted against its total experiment
    count. The plot includes only studies with at least one
    matching experiment ($n = 22{,}096$). A ratio of 1 indicates
    that all experiments in the study matched; the green
    curve ($1/N$, with $N$ denoting the number of experiments)
    identifies studies with exactly one matching experiment.
    Intermediate points represent studies in which several,
    but not all, experiments matched. The horizontal axis is
    logarithmic. The term ``annotated'' on the vertical axis
    refers to keyword detection under the original criteria.}
    \label{fig:annotation-proportion-2026}
\end{figure}

Figure~\ref{fig:annotation-proportion-2026} reproduces
the form of Figure~3 in the original 2015 investigation
\cite{alnasir2015annotation}, plotting the proportion
of experiments matching all three historical keyword
groups against the number of experiments per study.
The reference curve identifies studies with exactly
one matching experiment. Most of the 22,096 keyword-positive studies have
high within-study coverage, although in some studies
only a subset of experiments contains matching terms. The proportion of single-experiment
studies must be considered when interpreting the
concentration of points at a ratio of one.

Equivalent binned counts were not reported for Figure~3
in 2015, preventing a numerical comparison of
within-study consistency from these summaries alone.

\subsection{Database integrity and submission-date coverage}
\label{subsec:limitations}
Relational checks identified 55,917 experiments
(0.134\%) without a corresponding \texttt{study} row;
the study--experiment join contained 711,671 of
756,894 study accessions. These non-joins alone
do not demonstrate that source records are missing.
Of 7,891,372 submission records, 7,846,951
(99.44\%) had NULL \texttt{submission\_date} values.
The independent source-XML check reproduced this
date-availability count without discrepancies.
The limited date coverage precluded an archive-wide
analysis of annotation by submission year. Detailed
relational checks, temporal queries and incomplete
historical URL-query outputs are documented in the
Supplementary Information.

\section{Discussion}

\subsection{Changes in protocol keyword coverage since 2015}
The SRA contained almost 100 times as many experiments in September 2026 as in the December 2013 database. Over the same interval, the proportion of experiments matching all three historical keyword groups in \texttt{library\_construction\_protocol} fell from 4.06\% to 1.42\%. The number of matching records increased 34.8-fold, well below the growth of the experiment population. The greater frequency of NULL protocol fields also contributes to the reduction in archive-wide keyword coverage.

The results differ according to the metadata field examined.
In 2013, keyword matches were more frequent in
\texttt{library\_construction\_protocol} than in
\texttt{design\_description} for each of the three groups.
The order is reversed in 2026. The experiment-level
comparison quantifies the consequence for metadata
retrieval: a search confined to the designated field
misses 50.60\% of experiments that meet the historical
three-step keyword criterion when both fields are
considered. The overlap is small, and almost all
experiments matching all three groups exclusively in the
design description have a NULL protocol field.
Thus, the choice of metadata field materially affects
retrieval under the original analytical definition.

These findings concern where terms are detected.
A keyword in \texttt{design\_description} may describe
the experimental objective or another part of the
workflow, and its presence does not establish that a
complete preparation protocol has been supplied. The
snapshot comparison cannot determine whether individual
submitters have changed their choice of metadata field
since 2013.

\subsection{Variation in annotation within studies}
A large majority of contemporary studies joined to experiment records had no detectable historical keywords in the designated library-construction field. Among studies with at least one experiment matching all three groups, however, 95.30\% had matches in more than 75\% of their experiments. These figures refer to different populations and describe the distribution of keyword matches at different levels of the archive.

The original study reported marked variation in annotation among experiments belonging to the same study \cite{alnasir2015annotation}. The updated scatter plot (Figure~\ref{fig:annotation-proportion-2026}) likewise contains studies with only a small proportion of matching experiments. The concentration near complete within-study coverage needs to be examined against study size. A study containing a single experiment is assigned 100\% coverage whenever that experiment matches. Study-size-stratified results would allow a firmer comparison with the variation reported in 2015.

\subsection{Limitations of the historical keyword lists}
The unchanged SQL queries make the historical measurement repeatable on the reconstructed database. Their vocabulary was assembled for sequencing practices prevalent in 2013. Modern library preparation may use transposases to fragment DNA and attach adapters in a combined operation; other protocols avoid amplification altogether. For those workflows, the three historical categories may fail to recognise an informative description or may describe steps that were never applicable.

The sequencing-strategy analysis provides direct evidence of heterogeneity in the contemporary archive (Table~\ref{tab:strategy-coverage}). In particular, the collective three-step match rate was 0.26\% for \texttt{AMPLICON}, which accounted for 43.62\% of experiments, compared with 7.69\% for \texttt{RNA-Seq}. The choice of text field also had markedly different consequences for different strategies. These observations make it inappropriate to interpret the aggregate fall in historical keyword coverage as a direct measure of deteriorating metadata quality. They do not establish how much of the change since 2013 is attributable to shifts in the composition of sequencing strategies, because comparable historical strata have not been analysed.

Keyword detection has further limitations that apply to both snapshots. A protocol may explicitly state that PCR was omitted, refer only to a commercial kit, or mention a procedure without specifying the experimental conditions. Each can contain a historical search term while conveying very different information. The results measure the occurrence of those terms in selected fields. The complete agreement between XML-derived and SQLite-derived classifications in the validation sample addresses transcription and field-mapping consistency under these predicates; it does not establish semantic precision, sensitivity or the adequacy of the underlying laboratory documentation.

A separate contemporary vocabulary, with terms assigned according to the procedures they identify, is needed to estimate how much information the original lists miss. It should distinguish methods that replace a historical step from statements that a step was deliberately omitted. Applying both vocabularies within comparable sequencing strategies would help separate changes in terminology from changes in annotation practice. A manually reviewed sample could then provide estimates of false-positive and false-negative keyword matches, which were not measured in 2015.

\subsection{Reproducibility and metadata provenance}
The deposited SQL queries were applied to a new database constructed directly from NCBI XML while retaining the original SRAdb schema. This establishes that the historical analytical definitions can be used with a much larger contemporary archive. Exact reproduction of the 2015 published figures against the 2013 database remains a separate task. The original publication contains different experiment denominators and all-three-step counts in its narrative and Table~3; some supplementary scripts also use different aggregation definitions. These differences have been retained in the historical record and distinguished from the new measurements.

The continued usability of the original analytical scripts more
than a decade after publication is itself a noteworthy outcome
of this investigation. Their successful reuse against a substantially larger database containing almost one hundred times as many experiment records illustrates the enduring scientific value of preserving executable research methods alongside published findings. This experience provides a practical demonstration of how openly available analytical code and compatible data structures can support reproducible research over extended periods.

The source-level submission-date audit also showed that an incomplete administrative field cannot reliably support an archive-wide temporal analysis. The \texttt{submission\_date} attribute is optional and represents a submitter-assigned preparation date rather than necessarily the archive receipt date \cite{insdcSraSubmissionSchema}. The provenance and historical qualifications of this finding are given in the Supplementary Information.

\subsection{Implications for the reuse of sequencing data}
The 2015 investigation recommended more consistent recording of library preparation methods and greater use of controlled vocabularies \cite{alnasir2015annotation}. The contemporary results give reason to revisit those recommendations, particularly the distinction between a designated protocol field and information scattered through other descriptions. Stevens et al. also recommend planning metadata collection during experimental design and checking the resulting annotations \cite{stevens2020rules}.

Submission systems should permit explicit statements that a procedure was omitted or was inapplicable. When a method was used, records should identify it in sufficient detail, or link to a stable and versioned protocol. A mandatory free-text field may still be completed with information that is of little value for subsequent analysis \cite{brazma2009miame}. Better identification of library preparation methods would improve the prospects for comparisons across the increasingly heterogeneous SRA. The present results establish the scale of the differences detected by the historical vocabulary; their relationship to the quality of contemporary protocol descriptions requires workflow-specific evaluation.

\section{Conclusions}

Reapplying the original 2015 SQL keyword criteria to the September 2026 SRA metadata revealed a substantial change in the proportion of experiments containing detectable library preparation terminology. Matches to all three historical keyword groups in \texttt{library\_construction\_protocol} declined from 4.06\% in December 2013 to 1.42\% in September 2026, despite considerable growth in the number of matching records. This reduction measures the coverage of a fixed historical vocabulary within an expanding and increasingly heterogeneous archive. It does not, in itself, establish a deterioration in the quality of experimental documentation.

The principal contemporary finding concerns the distribution of historical keyword matches between metadata fields. Searching \texttt{design\_description} alongside the designated library construction field increased the number of experiments satisfying the three-step criterion from 591,526 to 1,197,367 (1.42\% to 2.88\%). A search restricted to the designated field would therefore miss 605,841 (50.60\%) of the experiments detectable by searching both fields collectively. Of the 605,757 experiments meeting the criterion exclusively in the design description, 604,116 had a NULL library construction protocol field. This quantifies a substantial limitation of field-restricted keyword retrieval, without implying that the additional matches represent complete or adequate laboratory protocols.

The pronounced differences observed between sequencing library strategies further demonstrate the limitations of applying historical keyword definitions indiscriminately to contemporary data. As preparation methods evolve, annotation practices must accommodate procedures that have been replaced, combined or deliberately omitted. Consistent use of designated metadata fields, supported by sufficiently detailed descriptions and stable protocol references, would improve the interpretability and reuse of archived experiments. Assessing the completeness of contemporary protocol documentation will nevertheless require workflow-specific terminology and independent examination of the information conveyed by individual records.

\section*{Availability of Data and Materials}

The historical SQL scripts and associated deposited dataset are
available through the original \emph{GigaScience} archive
\cite{alnasir2015software}. The September 2026 snapshot was
drawn from the NCBI full metadata archive identified above.

The additional SQL queries, database reconstruction scripts and
validation materials developed for the present investigation are
being prepared for public release. Details of their availability
will be provided in a subsequent version of this preprint.

\section*{Author Contributions}
Dr. Jamie J. Alnasir conceived and designed the follow-up investigation,
acquired and reconstructed the contemporary SRA metadata, developed
and executed the analyses and source-XML validation, interpreted the
results, and wrote and revised the manuscript.

\section*{Funding}
\noindent\emph{The author received no specific funding for this work.}

\section*{Competing Interests}
\noindent\emph{The author declares no competing interests.}

\section*{Acknowledgements}
I am sincerely grateful to Professor Hugh P. Shanahan
(Royal Holloway, University of London) for his mentorship
and collaboration during the original 2015
investigation. His thoughtful guidance helped
shape that research, which provided the foundation for
the questions revisited in the present study.


\begin{thebibliography}{99}

\bibitem{alnasir2015annotation}
Alnasir J, Shanahan HP.
Investigation into the annotation of protocol sequencing steps in the sequence read archive.
\emph{GigaScience}. 2015;4:23.
doi:10.1186/s13742-015-0064-7.

\bibitem{stevens2020rules}
Stevens I, Mukarram AK, H\"ortenhuber M, Meehan TF, Rung J, Daub CO.
Ten simple rules for annotating sequencing experiments.
\emph{PLOS Computational Biology}. 2020;16(10):e1008260.
doi:10.1371/journal.pcbi.1008260.

\bibitem{zhu2013sradb}
Zhu Y, Stephens RM, Meltzer PS, Davis SR.
SRAdb: query and use public next-generation sequencing data from within R.
\emph{BMC Bioinformatics}. 2013;14:19.
doi:10.1186/1471-2105-14-19.

\bibitem{katz2022sra}
Katz K, Shutov O, Lapoint R, Kimelman M, Brister JR, O'Sullivan C.
The Sequence Read Archive: a decade more of explosive growth.
\emph{Nucleic Acids Research}. 2022;50(D1):D387--D390.
doi:10.1093/nar/gkab1053.

\bibitem{ross2013bias}
Ross MG, Russ C, Costello M, Hollinger A, Lennon NJ, Hegarty R, et al.
Characterizing and measuring bias in sequence data.
\emph{Genome Biology}. 2013;14:R51.
doi:10.1186/gb-2013-14-5-r51.

\bibitem{kozarewa2009amplification}
Kozarewa I, Ning Z, Quail MA, Sanders MJ, Berriman M, Turner DJ.
Amplification-free Illumina sequencing-library preparation facilitates improved mapping and assembly of (G+C)-biased genomes.
\emph{Nature Methods}. 2009;6:291--295.
doi:10.1038/nmeth.1311.

\bibitem{klie2021metadata}
Klie A, Tsui BY, Mollah S, Skola D, Dow M, Hsu C-N, Carter H.
Increasing metadata coverage of SRA BioSample entries using deep learning-based named entity recognition.
\emph{Database}. 2021;2021:baab021.
doi:10.1093/database/baab021.

\bibitem{wilkinson2016fair}
Wilkinson MD, Dumontier M, Aalbersberg IJ, Appleton G, Axton M, Baak A, et al.
The FAIR Guiding Principles for scientific data management and stewardship.
\emph{Scientific Data}. 2016;3:160018.
doi:10.1038/sdata.2016.18.

\bibitem{ncbiSraMetadata}
National Center for Biotechnology Information.
SRA Metadata and Submission Overview.
\url{https://www.ncbi.nlm.nih.gov/sra/docs/submitmeta/}
(accessed 11 October 2026).

\bibitem{ncbiBioSample}
National Center for Biotechnology Information.
BioSample Documentation.
\url{https://www.ncbi.nlm.nih.gov/biosample/docs/}
(accessed 11 October 2026).

\bibitem{alnasir2015software}
Alnasir J, Shanahan HP.
Supporting material for ``Investigation into the annotation of protocol sequencing steps in the Sequence Read Archive''.
\emph{GigaScience Database}. 2015.
\url{https://doi.org/10.5524/100142};
\url{https://github.com/gigascience/paper-alnasir2015}.

\bibitem{brazma2009miame}
Brazma A.
Minimum Information About a Microarray Experiment (MIAME) --- successes, failures, challenges.
\emph{The Scientific World Journal}. 2009;9:420--423.
doi:10.1100/tsw.2009.57.

\bibitem{rayner2006magetab}
Rayner TF, Rocca-Serra P, Spellman PT, Causton HC, Farne A, Holloway E, et al.
A simple spreadsheet-based, MIAME-supportive format for microarray data: MAGE-TAB.
\emph{BMC Bioinformatics}. 2006;7:489.
doi:10.1186/1471-2105-7-489.

\bibitem{nekrutenko2012reproducibility}
Nekrutenko A, Taylor J.
Next-generation sequencing data interpretation: enhancing reproducibility and accessibility.
\emph{Nature Reviews Genetics}. 2012;13:667--672.
doi:10.1038/nrg3305.

\bibitem{insdcSraSubmissionSchema}
International Nucleotide Sequence Database Collaboration.
SRA Submission XML specification, version 1.5.9
(\texttt{SRA.submission.xsd}). DDBJ public schema repository.
\url{https://github.com/ddbj/pub/blob/master/docs/dra/xsd/1-5/SRA.submission.xsd}
(accessed 14 October 2026).

\bibitem{seqanswers2014date}
Alnasir J. SRA top level studies counts, why is 2013 so low?
\emph{SEQanswers}. June 2014. Public discussion containing
a response attributed to an SRAdb developer on missing dates.
\url{https://www.seqanswers.com/forum/bioinformatics/bioinformatics-aa/38501-sra-top-level-studies-counts-why-is-2013-so-low}
(accessed 14 October 2026).


\end{thebibliography}
\end{document}